\UseRawInputEncoding
\documentclass[12pt,letter]{article}
\pdfoutput=1
\usepackage{graphicx, epsfig, color,cite}
\usepackage{amsmath}
\usepackage{amssymb}
\usepackage{float}
\usepackage{caption,subcaption,graphicx}
\usepackage{hyperref}

\def\bi{\begin{itemize}}
\def\ei{\end{itemize}}

\def\tu{\tilde u}

\def\tb{\tilde b}

\def\tst{\tilde t}

\def\tg{\tilde g}

\def\tz{\widetilde\chi^0}
\def\alt{\lesssim}
\def\agt{\gtrsim}
\def\be{\begin{equation}}  
\def\ee{\end{equation}}  
\def\bea{\begin{eqnarray}}  
\def\eea{\end{eqnarray}}

\begin{document}
\begin{titlepage}
\begin{flushright}
OU-HEP-201104
\end{flushright}

\vspace{0.5cm}
\begin{center}
{\Large \bf Landscape Higgs and sparticle mass predictions\\
from a logarithmic soft term distribution
}\\ 
\vspace{1.2cm} \renewcommand{\thefootnote}{\fnsymbol{footnote}}
{\large Howard Baer$^1$\footnote[1]{Email: baer@ou.edu },
Vernon Barger$^2$\footnote[2]{Email: barger@pheno.wisc.edu},
Shadman Salam$^1$\footnote[3]{Email: shadman.salam@ou.edu} \\
and
Dibyashree Sengupta$^3$\footnote[4]{Email: Dibyashree.Sengupta-1@ou.edu}
}\\ 
\vspace{1.2cm} \renewcommand{\thefootnote}{\arabic{footnote}}
{\it 
$^1$Homer L. Dodge Department of Physics and Astronomy,
University of Oklahoma, Norman, OK 73019, USA \\[3pt]
}
{\it 
$^2$Department of Physics,
University of Wisconsin, Madison, WI 53706 USA \\[3pt]
}
{\it 
$^3$Department of Physics,
National Taiwan University, Taipei 10617, Taiwan, R.O.C. \\[3pt]
}

\end{center}

\vspace{0.5cm}
\begin{abstract}
\noindent
Recent work on calculating string theory landscape statistical predictions 
for the Higgs and sparticle mass spectrum from an assumed power-law 
soft term distribution yields an expectation for $m_h\simeq 125$ GeV
with sparticles (save light higgsinos) somewhat beyond reach of 
high-luminosity LHC. 
A recent examination of statistics of SUSY breaking in
IIB string models with stabilized moduli suggests a power-law for
models based on KKLT stabilization and uplifting while models 
based on large-volume scenario (LVS) instead yield an expected 
logarithmic soft term distribution. We evaluate statistical distributions
for Higgs and sparticle masses from the landscape with a log soft term
distribution and find the Higgs mass still peaks around $\sim 125$ GeV
with sparticles beyond LHC reach, albeit with somewhat softer 
distributions than those arising from a power-law.

\end{abstract}
\end{titlepage}

\section{Introduction}
\label{sec:intro}

The cosmological constant problem-- how can it be that the numerical value
of $\Lambda_{cc}$ is more than 120 orders of magnitude less than its expected 
theoretical value\cite{Weinberg:1987dv2}-- 
finds a compelling resolution within the landscape of string 
vacua\cite{Susskind:2003kw} 
coupled to anthropic reasoning\cite{Bousso:2000xa}. The idea is that, with
say $\sim 10^{500}$ string vacua states\cite{Denef:2004dm,Douglas:2006es}, 
each leading to different $4-d$ laws 
of physics and with a rather uniform distrbutions of $\Lambda_{cc}$ ranging
from $m_P^2$ to values well below $10^{-120}m_P^2$, then it may not be surprising 
to find ourselves living within a pocket-universe (within the eternally-inflating multiverse) 
with such a small $\Lambda_{cc}$ since if its value was much larger, 
then the expansion rate would be so great that galaxy (structure) formation would not 
occur\cite{Weinberg:1987dv}. 
Such anthropic reasoning works if we posit in addition that our pocket universe
is but one within a fertile/friendly patch wherein the Standard Model (SM) 
is the low energy effective field theory (EFT) and only one of a few fundamental parameters 
(such as $\Lambda_{cc}$) scans within the patch\cite{Weinberg:2005fh}.

Possibly related to the cosmological constant problem is the riddle as to why-- 
within the SM-- the magnitude of the weak scale $m_{weak}\sim m_{W,Z,h}\sim 100$ GeV is 
so much lower than than the Planck or GUT scale. 
Even if tree level SM parameters are dialed so that $m_{weak}\sim 100$ GeV, quadratically divergent
quantum corrections to $m_h$ ought to drag its mass up to the highest energy scales admissible 
within the model. 
The latter situation is fixed\cite{witten,kaul} by supersymmetrizing the SM into the 
softly broken Minimal Supersymmetric Standard Model or MSSM\cite{Dimopoulos:1981zb,WSS}.
While weak-scale SUSY stabilizes the weak scale, it still doesn't explain its magnitude.

An explanation for the magnitude of the weak scale can also be found within the string
landscape wherein something like the MSSM forms the low energy $4-d$ EFT. 
In this case, one again assumes a fertile patch of vacua wherein just the soft SUSY breaking terms scan
between different pocket universes\cite{Douglas:2003um}. 
Then, the pocket-universe value for the weak scale is determined by the values of the
soft SUSY breaking terms and the superpotential $\mu$ term. While the former are expected to scan, 
the latter value is presumably determined by whatever solution to the SUSY $\mu$ problem
is invoked in nature\cite{Bae:2019dgg}. 
Then, the pocket-universe value of the weak scale is given (in terms of the $Z$-mass) by 
\be
\frac{(m_Z^{PU})^2}{2}=\frac{m_{H_d}^2+\Sigma_d^d-(m_{H_u}^2+\Sigma_u^u)\tan^2\beta}{\tan^2\beta -1}-\mu^2\simeq -m_{H_u}^2-\Sigma_u^u-\mu^2 .
\label{eq:mzs}
\ee
Here, $m_{H_u}^2$ and $m_{H_d}^2$ are the Higgs field soft squared-masses and the
$\Sigma_d^d$ and $\Sigma_u^u$ contain over 40 loop corrections to the weak scale
(expressions can be found in the Appendix to Ref. \cite{rns})). 
Agrawal {\it et al.}\cite{Agrawal:1997gf,Agrawal:1997gf2} have calculated that if the pocket-universe (PU) value $m_{weak}^{PU}$
is much bigger or smaller than its measured value in our universe (OU), then complex nuclei
and hence atoms as we know them would not form. 
The existence of atoms seems necessary for life as we know it; this is known as the
{\it atomic principle}, in analogy with the stucture principle used for anthropic reasoning
in the case of the cosmological constant.
Thus, a value of $m_{weak}^{PU}$ to within a factor of 2-5 of its value in our universe would form
the anthropic requirement for life as we know it in a fertile patch of vacua with an MSSM-like 
low energy EFT.

Douglas has proposed a program for the statistical determination of the soft SUSY breaking 
terms, and hence the expected Higgs and sparticle mass spectrum, from sampling the
array of soft terms expected within our fertile patch\cite{Douglas:2004qg}.
The distribution of $4-d$ supergravity vacua with hidden sector SUSY breaking scale $m_{hidden}^2$ 
is expected to be a product of three distributions
\be
dN_{vac}(m_{hidden}^2,m_{weak},\Lambda_{cc} )= f_{SUSY}\cdot f_{EWSB}\cdot f_{cc}\cdot dm_{hidden}^2
\label{eq:dNvac}
\ee
where the hidden sector SUSY breaking scale 
$m_{hidden}^4=\sum_i |F_i|^2 \, +  \, \frac{1}{2}\sum_\alpha D^2_\alpha $
is a mass scale associated with the hidden sector
(and usually in SUGRA-mediated models it is assumed $m_{hidden}\sim 10^{12}$ GeV
such that the gravitino gets a mass $m_{3/2}\sim m_{hidden}^2/m_P$).
Consequently, in gravity-mediation then the visible sector soft terms are of 
magnitude $m_{soft}\sim m_{3/2}$\cite{Soni:1983rm,Kaplunovsky:1993rd,Brignole:1993dj}. 

As noted by Susskind\cite{susskind} and Douglas\cite{denefdouglas}, 
the scanning of the cosmological constant is effectively independent 
of the determination of the SUSY breaking scale so that $f_{cc}\sim \Lambda_{cc} /m_{string}^4$. 
Thus, the cosmological constant decouples from the statistical determination
of the SUSY breaking scale.

One proposal for $f_{SUSY}$ arises from examining flux vacua in IIB string 
theory\cite{Douglas:2004qg,susskind,ArkaniHamed:2005yv}.
Since nothing in string theory prefers one SUSY breaking vev over another, then
all values should be comparably probable. Then the SUSY breaking $F_i$ and $D_\alpha$ terms
are likely to be uniformly distributed-- 
in the former case as complex numbers while in the latter case as real numbers.
Then one expects the following distribution of supersymmetry breaking scales
\be
f_{SUSY}(m_{hidden}^2) \, \sim \, (m^2_{hidden})^{2n_F+n_D - 1}
\label{eq:fSUSY}
\ee
where $n_F$ is the number of $F$-breaking fields and $n_D$ is the number of 
$D$-breaking fields in the hidden sector.
Even for the case of just a single $F$-breaking term, then one expects a {\it linear} 
statistical draw towards large soft terms; $f_{SUSY}\sim m_{soft}^n$
where $n=2n_F+n_D-1$ and in this case where $n_F=1$ and $n_D=0$ then $n=1$.
For SUSY breaking contributions from multiple hidden sectors, as typically
expected in string theory, then $n$ can be much larger, with a consequent 
stronger pull towards large soft breaking terms.

An initial guess for $f_{EWSB}$, the (anthropic) electroweak symmetry breaking factor,
was $m_{weak}^2/m_{soft}^2$ which would penalize soft terms which were much
bigger than the weak scale. However, it was pointed out in Ref's \cite{Baer:2016lpj,Baer:2017uvn}
that this ansatz fails in a variey of circumstances:
\begin{itemize}
\item If the trilinear soft breaking parameter $A_t$ gets too big, then the $m_{\tst_R}^2$
soft term can be driven negative resulting in charge or color breaking (CCB) vacua. 
Such vacua states would likely be hostile to life as we know it, and would have to be vetoed
instead of merely penalized.
\item If soft terms such as $m_{H_u}^2$ are too big, then they will not be driven negative at 
the weak scale and EW symmetry will not even break; such vacua should also be vetoed.
\item For some soft terms, the larger their high scale value, then the {\it smaller} 
(more natural) is their associated contribution to the weak scale. 
This occurs for $m_{H_u}^2$, where if it is too small, it is driven deeply negative to 
(negative) multi-TeV squared values. Then one would expect $m_{weak}^{PU}$ to also be in 
the multi-TeV range absent any (improbable) fine-tuning. Also, the larger the $A_t$ parameter, 
then the smaller are the $\Sigma_u^u(\tst_{1,2})$ contributions to the weak 
scale (due to cancellations)\cite{ltr,stringy}.
And also, large first/second generation soft terms $m_0^2(1,2)$ in the tens of TeV range can drive 
$m_{\tst_{L,R}}^2$ to small weak scale values via 2-loop RG effects; 
this would also result in smaller $\Sigma_u^u(\tst_{1,2})$ contributions to the weak scale.
\end{itemize}
To ameliorate these issues, the form
\be
f_{EWSB}=\Theta (30 -\Delta_{EW})
\label{eq:fewsb}
\ee
was adopted, which requires all contributions to the weak scale to be within a factor 
four of the measured value of the weak scale in our universe $m_{weak}^{OU}$. 
In addition, EW symmetry must be properly broken with no CCB minima.

The statistical calculation of generating soft terms according to $f_{SUSY}$ (as might be expected in a scan
of our fertile patch of the landscape) along with the anthropic selection Eq. \ref{eq:fewsb} that
EWSB is properly broken with $m_{weak}^{PU}\alt 4 m_{weak}^{OU}$, has led to some very compelling success.
The statistical pull on soft SUSY breaking terms to large values pulls the sparticle mass predictions beyond 
present day limits from LHC: for instance, statistically we would expect for $n=1$ or 2 that
$m_{\tg}\sim 4\pm 2$ TeV which explains why LHC hasn't seen gluinos so far. 
But sparticles can't be pulled so high that $m_{weak}^{PU}$ gets too big-- lest we violate the 
atomic principle. The fact that $A_t$ gets pulled to large values (but stopping short of CCB vacua)
leads to maximal mixing in the top-squark sector which also lifts $m_h$ to a statistical peak
around $m_h\simeq 125$ GeV. Thus, from the string landscape with a power-law draw to large soft terms, 
we expect $m_h\simeq 125$ GeV while sparticles are typically beyond present LHC reach. 
This methodology has been applied to the two- and three-extra parameter non-universal Higgs 
gravity-mediation models\cite{nuhm2,nuhm22,nuhm23,nuhm24,nuhm25,nuhm26} NUHM2 and NUHM3 in Ref. \cite{Baer:2017uvn} 
and to mirage mediation in Ref. \cite{Baer:2019tee}.

\begin{figure}[tbh]
\begin{center}
\includegraphics[height=0.4\textheight]{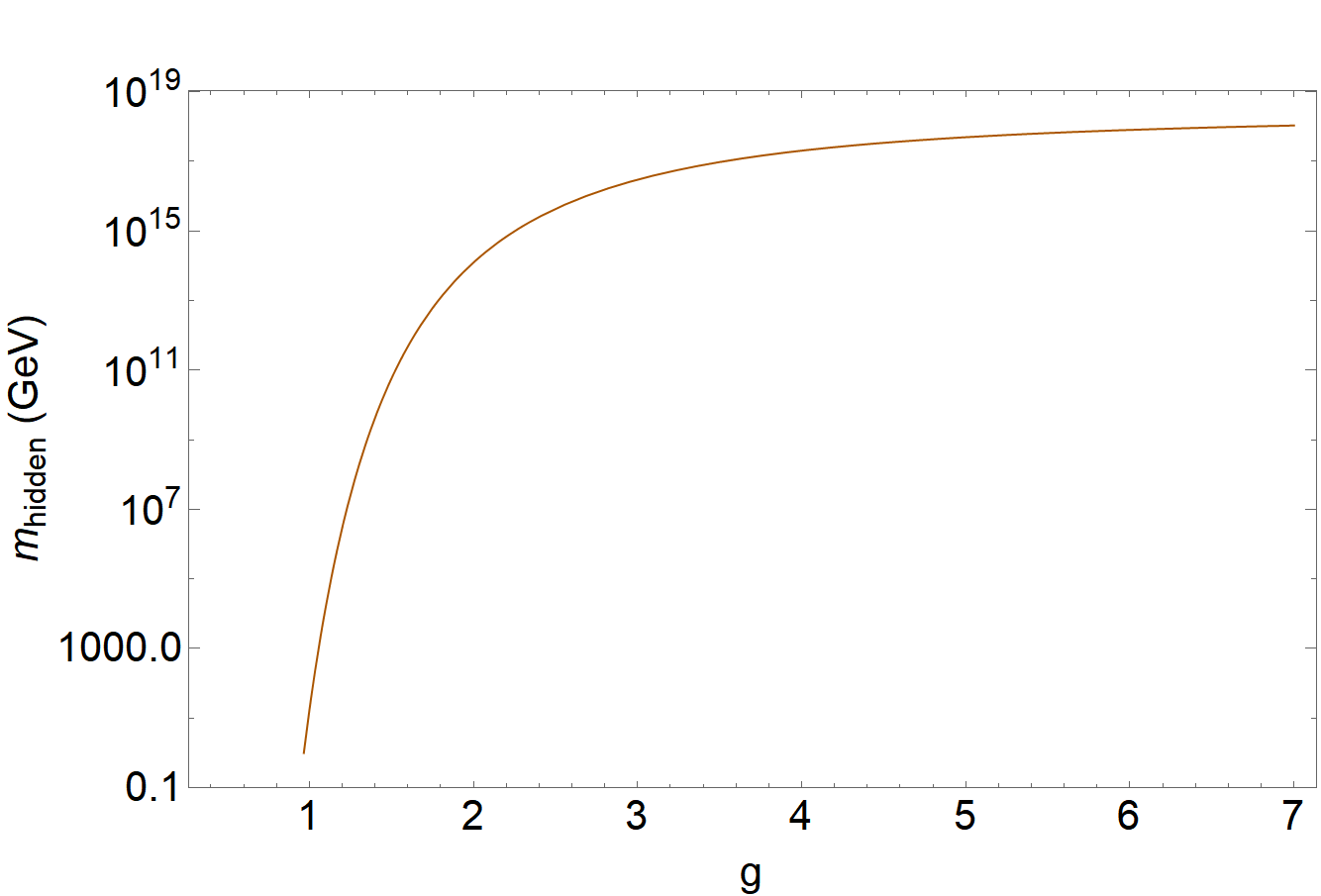}
\caption{Expected SUSY breaking scale $m_{hidden}$ vs. 
hidden sector coupling $g$ from dynamical SUSY breaking.
\label{fig:mvsg}}
\end{center}
\end{figure}

An alternative form for $f_{SUSY}$ is instead motivated by {dynamical SUSY breaking} (DSB) 
where instead of perturbative breaking, 
one expects SUSY breaking from non-perturbative effects\cite{witten}.
These might include for instance SUSY breaking via gaugino condensation when some hidden sector
gauge group becomes strongly interacting. In such a case, the hidden sector SUSY breaking scale is
expected to be $m_{hidden}^2\sim m_P^2 e^{-8\pi^2/b_0 g^2}$ where $g$ is the hidden sector gauge coupling
and $b_0\sim 1$ enters the hidden sector beta function. A strong motivation for DSB is that it 
provides some mechanism to generate an exponential hierarchy between the SUSY breaking scale and 
the fundamental scale $m_P$ in the theory. 
A plot of hidden sector SUSY breaking mass scale vs. $g_{hidden}$ is shown in Fig. \ref{fig:mvsg}.
In the context of the landscape, if the hidden gauge coupling $g_{hidden}^2$ is uniformly 
distributed within various pocket universes\cite{denefdouglas}, 
then the SUSY breaking scale will be logarithmically distributed across the decades of values. 
Then we might expect as well a slowly rising (log) distribution of soft terms. 
A log or log times a power law  distribution of soft terms has been advanced by Dine {\it et al.}
in a series of papers\cite{Dine:2004is,Dine:2004ct,Dine:2005yq,Dine:2005iw}. Such a log distribution may also be expected in non-perturbative 
SUSY breaking due to instanton effects\cite{Affleck:1983rr}.

The logarithmic landscape draw on soft terms has also emerged from considerations 
of K\"ahler moduli stabilization in Ref. \cite{Broeckel:2020fdz}. 
In previous work, K\"ahler moduli effects were thought to be subleading, 
but in Ref. \cite{Broeckel:2020fdz}
the stabilization of K\"ahler moduli in KKLT\cite{Kachru:2003aw}, 
LVS\cite{Balasubramanian:2005zx} and perturbative moduli stabilization\cite{Berg:2005yu} (PS) 
schemes was examined. 
For these cases, it was found that KKLT and PS led to a power-law draw 
for soft terms whilst LVS stabilization led to a log draw.
Given these motivations, in the present paper we calculate statistical distributions 
of Higgs and sparticle masses in the NUHM3 gravity-mediation model 
assuming a log draw on soft terms.

\section{Results}
\label{sec:results}

In this Section, we will present the results of calculations of the string landscape
probability distributions for Higgs and sparticle masses under the assumption of 
$f_{SUSY}=\log(m_{soft})$ along with Eq. \ref{eq:fewsb} for $f_{EWSB}$. 
Our results will be presented within the gravity-mediated three extra parameter
non-universal Higgs model NUHM3 with parameter space given by
\be
m_0(1,2),\ m_0(3),\ m_{1/2},\ A_0,\ \tan\beta,\ \mu,\ m_A\ \ \ \ (NUHM3).
\ee
We adopt the Isajet\cite{isajet} code for calculation of Higgs and superparticle
mass spectrum\cite{Baer:1994nc} based on 2-loop RGE running\cite{Martin:1993zk} along with sparticle and Higgs masses
calculated at the RG-improved 1-loop level\cite{Pierce:1996zz}.

To compare our results against similar calculations which were presented in Ref. \cite{Baer:2017uvn}
but using $f_{SUSY}=m_{soft}^n$, we will scan over the same parameter space
\begin{itemize}
\item $m_0(1,2):\ 0.1 - 60$ TeV,
\item $m_0(3):\ 0.1 - 20$ TeV,
\item $m_{1/2}:\ 0.5 - 10$ TeV,
\item $A_0:\ -50 -\ 0$ TeV,
\item $m_A:\ 0.3 - 10$ TeV,
\end{itemize}
with $\mu = 150$ GeV while $\tan\beta:3-60$ is scanned uniformly.
The goal here is to choose upper limits to our scan parameters
which will lie beyond the upper limits imposed by the anthropic selection from $f_{EWFT}$.
Lower limits are motivated by current LHC search limits. 
Our final results will hardly depend on the chosen value of $\mu$ so long as 
$\mu$ is within an factor of a few of $m_{W,Z,h}\sim 100$ GeV.

\begin{figure}[H]
\begin{center}
\includegraphics[height=0.22\textheight]{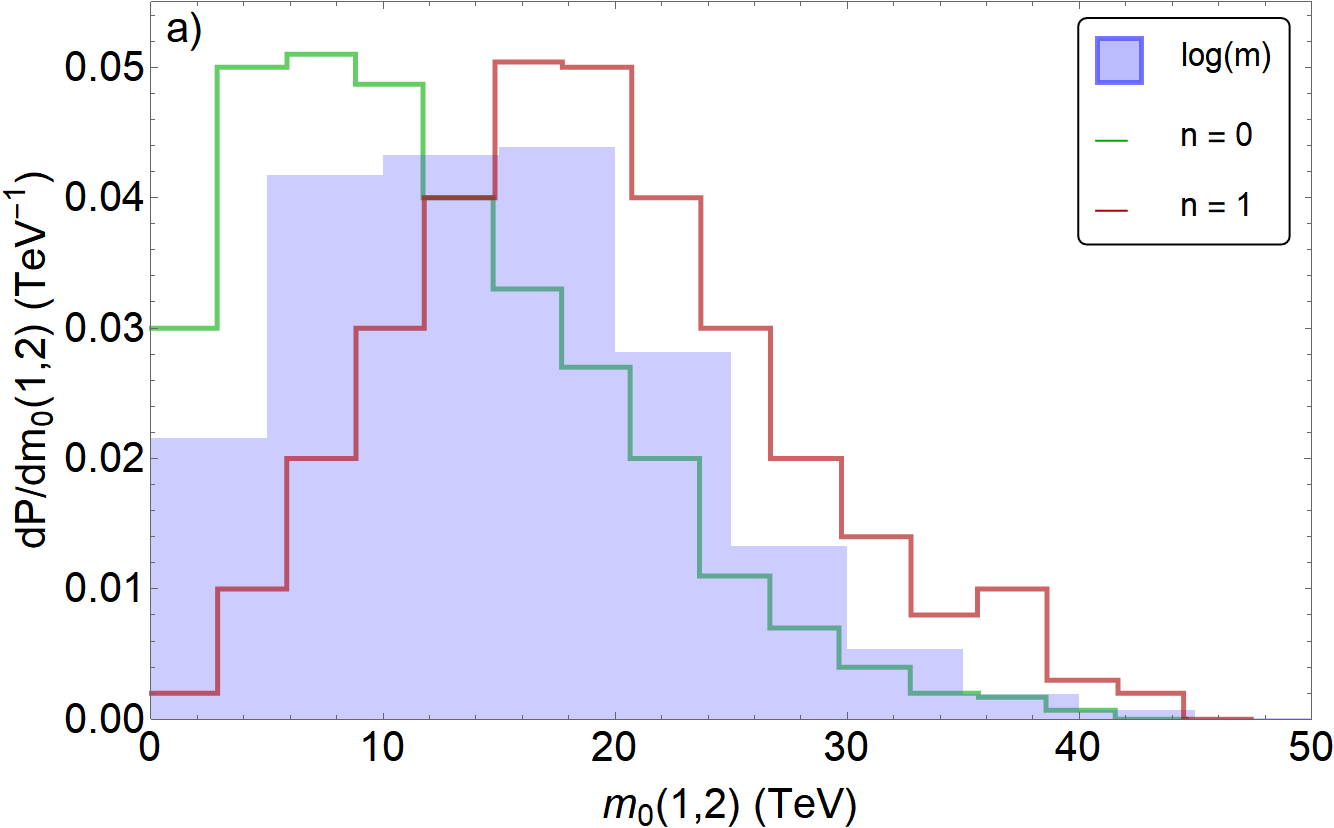}
\includegraphics[height=0.22\textheight]{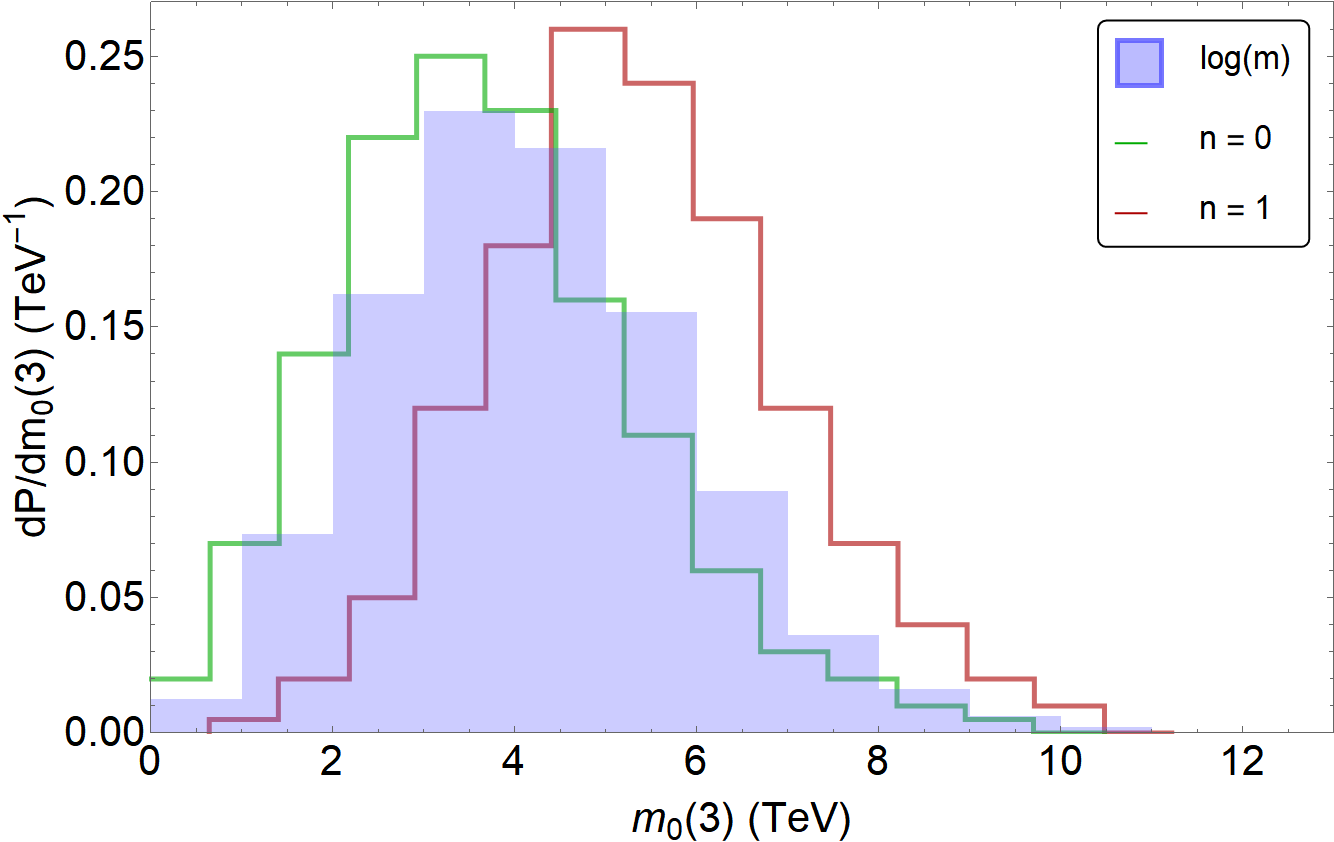}\\
\includegraphics[height=0.22\textheight]{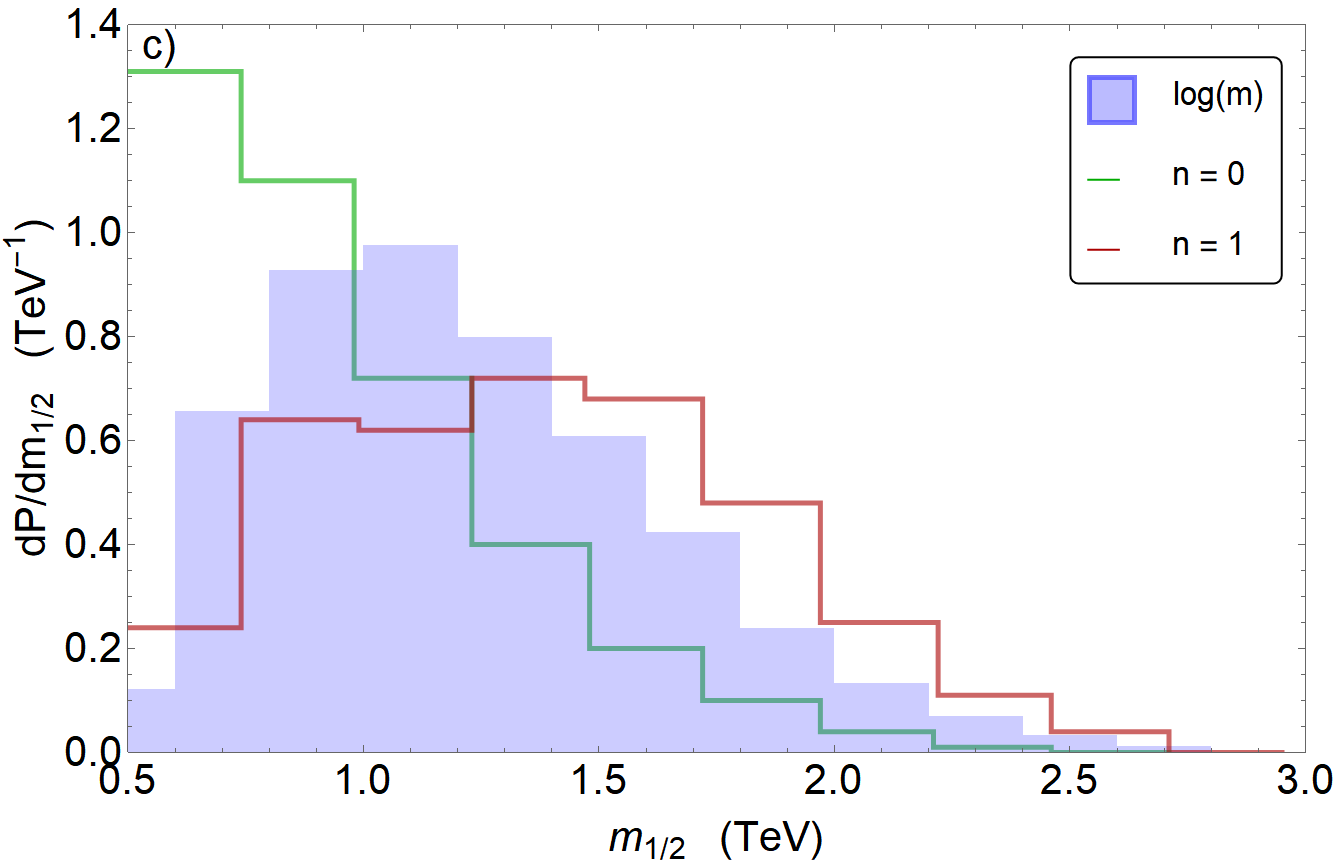}
\includegraphics[height=0.22\textheight]{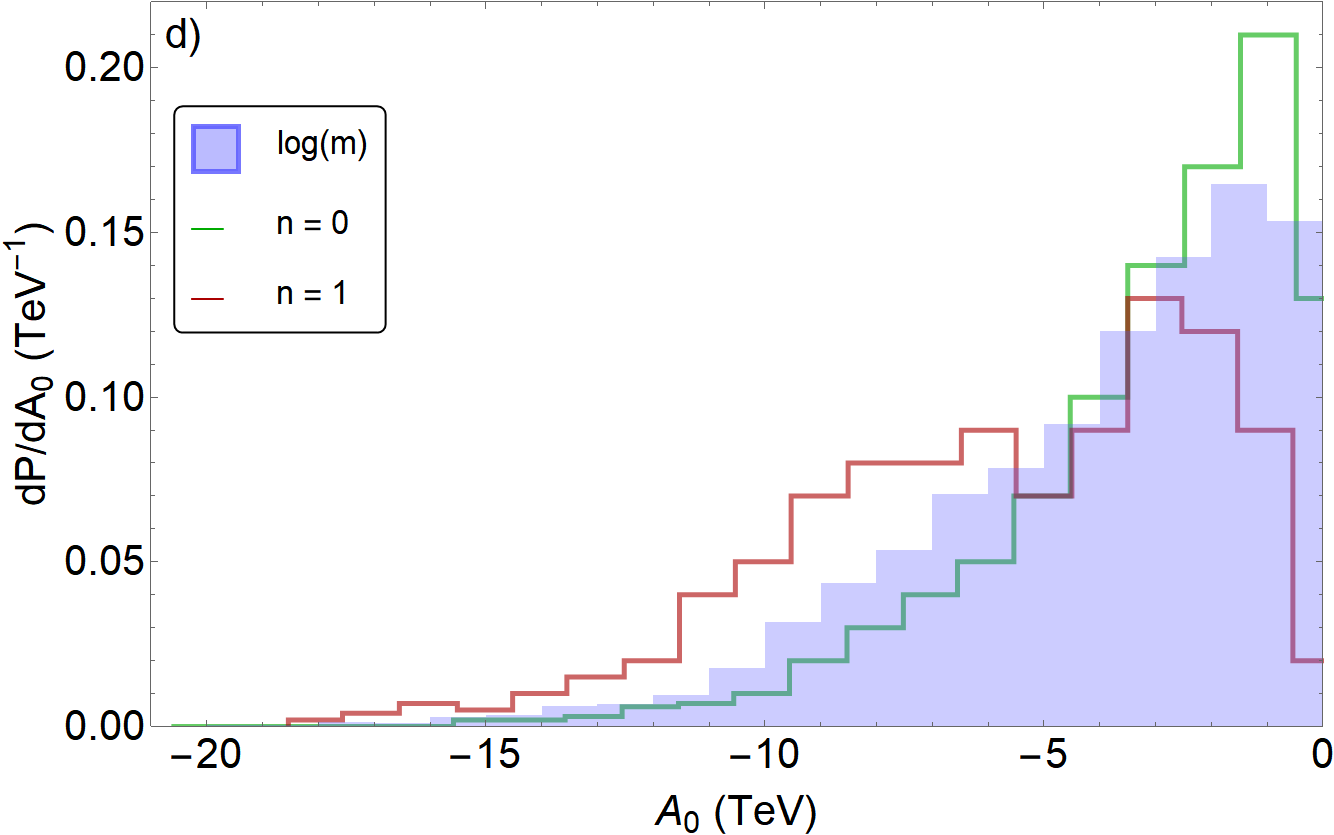}
\caption{Probability distributions for NUHM3 soft terms
{\it a}) $m_0(1,2)$, {\it b}) $m_0(3)$, {\it c}) $m_{1/2}$ and
{\it d}) $A_0$ from a log distribution of soft terms
in the string landscape with $\mu =150$ GeV.
For comparison, we also show probability distributions for $n=0$ and 1.
\label{fig:m0mhf}}
\end{center}
\end{figure}

Our first results are shown in Fig. \ref{fig:m0mhf}, where we show probability 
distributions of input parameters. Our present $f_{SUSY}=\log (m_{soft})$ distributions are
shown as shaded histograms. These are then compared to previous scans with
$f_{SUSY}=m_{soft}^n$ for $n=0$ (typical uniform scan) and the simplest power-law
scan with $n=1$. In Fig. \ref{fig:m0mhf}{\it a}), we show the distribution vs. 
first/second generation soft terms $m_0(1,2)$. While the $n=0$ result (green histogram)
peaks at the few TeV range, and the $n=1$ result (red histogram) peaks around
15-20 TeV, we see that the new log result is significantly harder than the uniform scan 
but slightly softer than $n=1$. Here, the peak probability value for $m_0(1,2)$
lies in the 10-20 TeV range with a tail extending to 40 TeV. 
Such large values of $m_0(1,2)$  should still be enough to provide the mixed
decoupling/quasi-degeneracy landscape solution to the SUSY flavor/CP problems as outlined in
Ref. \cite{Baer:2019zfl}. In frame {\it b}), we show the distribution vs. third generation 
soft mass $m_0(3)$. Here, the log distribution peaks around 4 TeV which is again
intermediate between the $n=0$ and $n=1$ histograms. For frame {\it c}), 
which shows the distribution in unified gaugino mass $m_{1/2}$, we see the log 
distribution peaks around 1.2 TeV with a tail extending to $\sim 2.5$ TeV. For 
comparison, the $n=0$ scan peaks at very low $m_{1/2}$ while the $n=1$ histogram peaks 
near $m_{1/2}\sim 1.5$ TeV. Finally, we show in frame {\it d}) the distribution in $-A_0$.
While the uniform scan favors small $|A_0|$, leading to small stop mixing and lower $m_h$, 
the log scan is somewhat harder: with repect to $n=0$, it has a lower pek
at small $|A_0|$ (leading to small stop mixing) and more of a bulge around $A_0\sim -7$ TeV 
(leading to large stop mixing).
The large $-A_0$ values provide previously noted {\it cancellations} in the
weak scale radiative corrections $\Sigma_u^u(\tst_{1,2})$ when $|A_0|$ becomes maximal
just stopping short of inducing CCB minima in the scalar potential\cite{ltr,stringy}.  

\begin{figure}[H]
\begin{center}
\includegraphics[height=0.22\textheight]{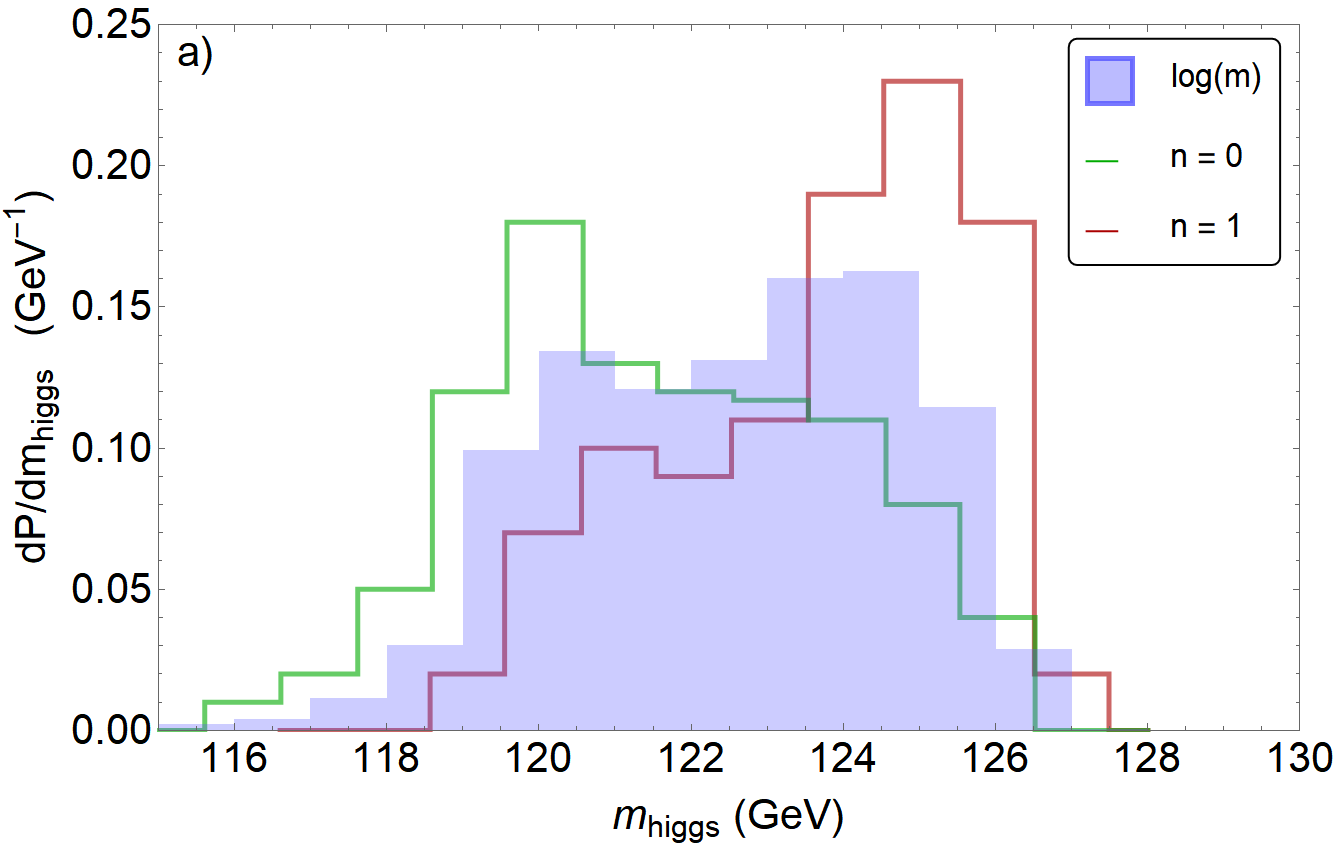}
\includegraphics[height=0.22\textheight]{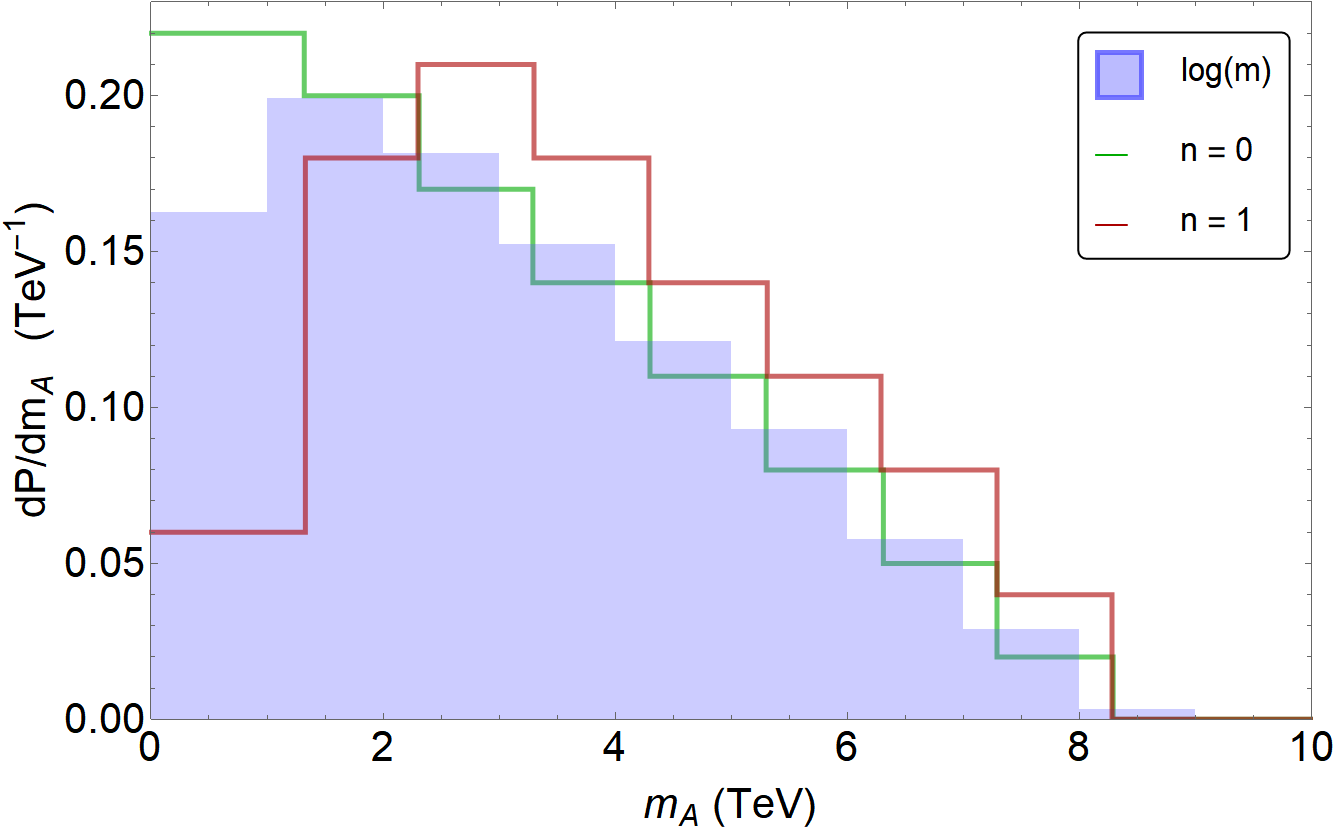}\\
\includegraphics[height=0.22\textheight]{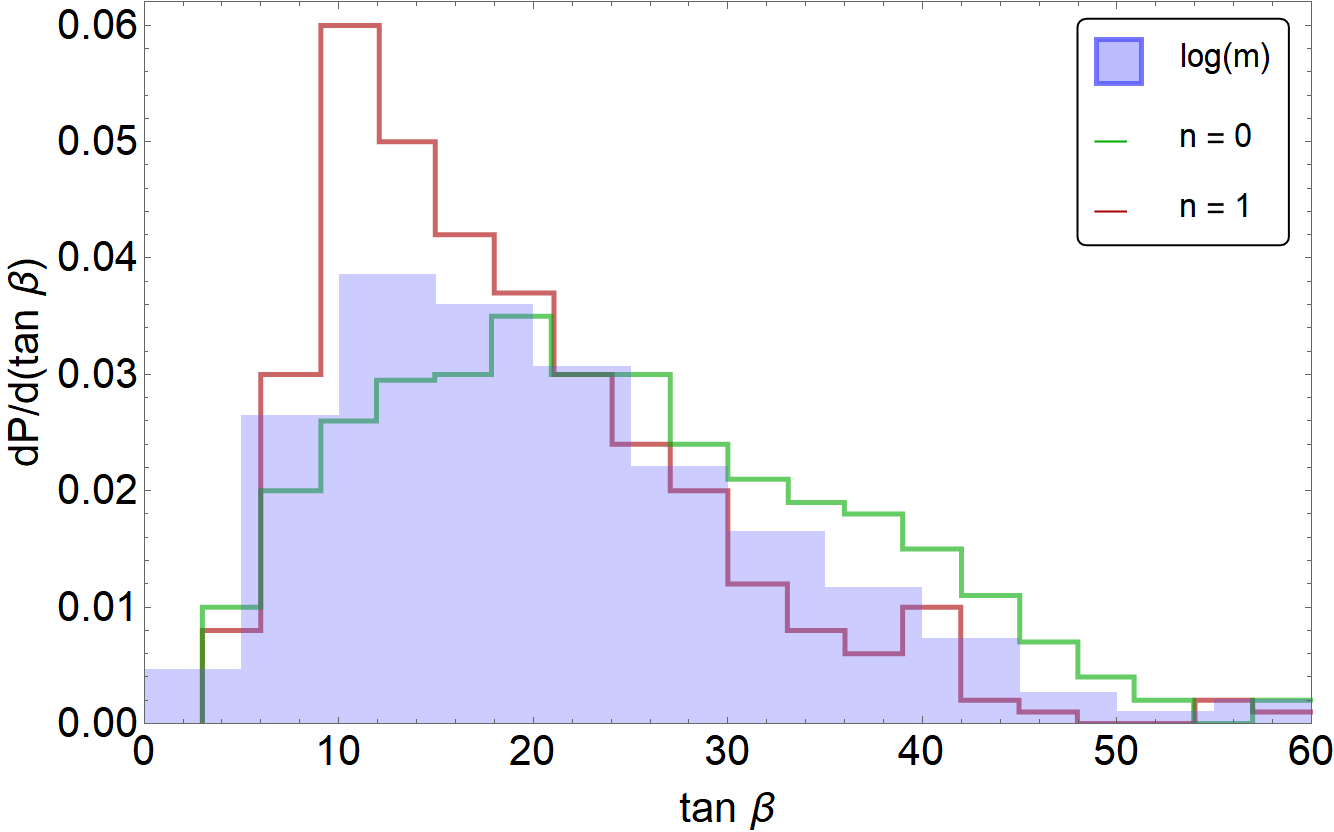}
\includegraphics[height=0.22\textheight]{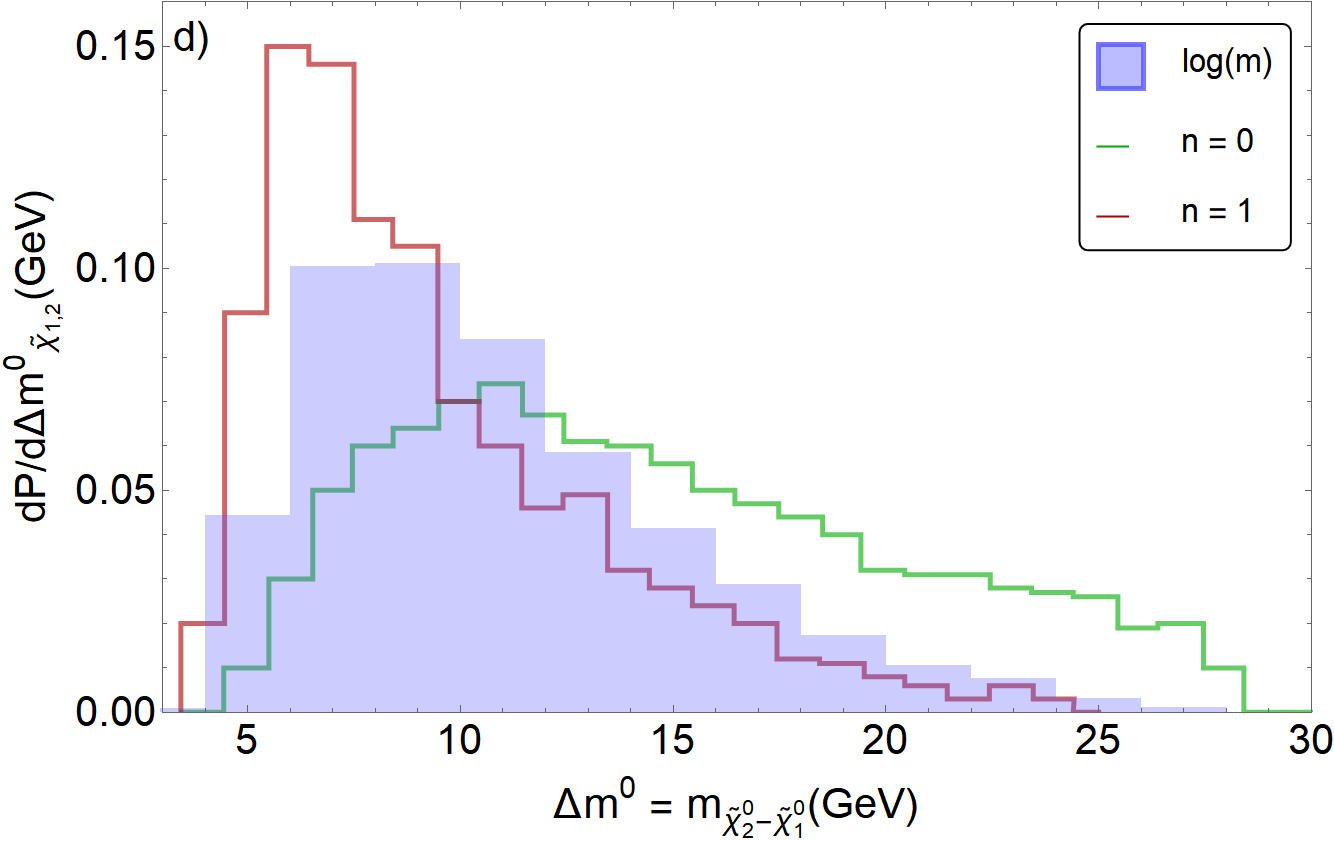}
\caption{Probability distributions for NUHM3 masses and 
parameters {\it a}) $m_h$, {\it b}) $m_A$, {\it c}) $\tan\beta$ and
{\it d}) $m_{\tz_2}-m_{\tz_1}$ from a log distribution of soft terms
in the string landscape with $\mu =150$ GeV.
For comparison, we also show probability distributions for $n=0$ and 1.
\label{fig:higgs}}
\end{center}
\end{figure}

\begin{figure}[H]
\begin{center}
\includegraphics[height=0.4\textheight]{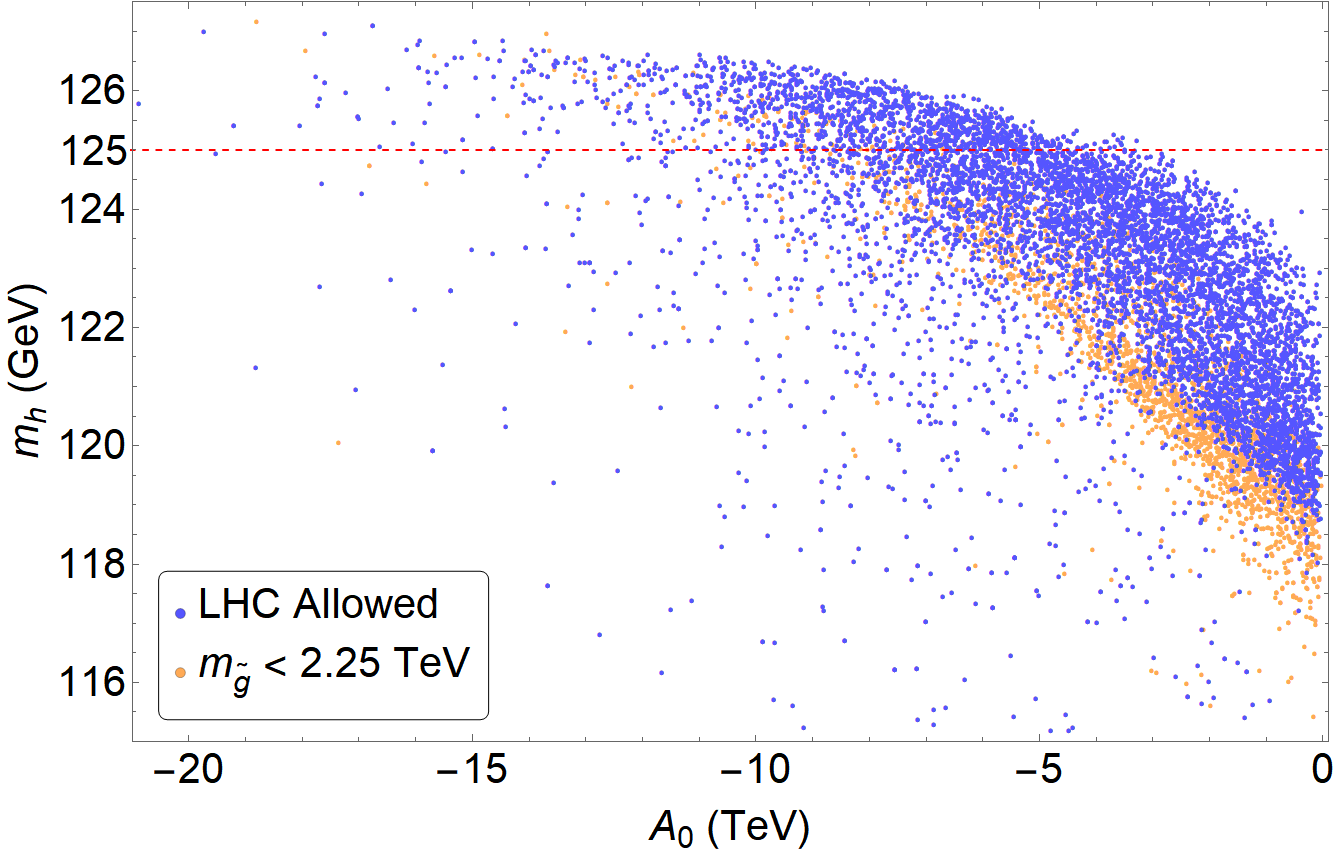}
\caption{Scatter plot of models with appropriate EWSB and $m_Z^{PU}<4m_Z^{OU}$
in the $m_h$ vs. $A_0$ plane with $\mu =150$ GeV.
\label{fig:mhvsA0}}
\end{center}
\end{figure}

In Fig. \ref{fig:higgs}, we show various probability distributions for
quantities associated with the Higgs/higgsino sector, including the light Higgs
mass $dP/dm_h$ in frame {\it a}). From the frame, we see that with a 
log distribution of soft terms, the distribution in $m_h$ almost has a two-peak
structure: a dominant peak around $m_h\sim 125$ GeV and a sub-dominant
peak around $m_h\sim 120$ GeV. The dominant large $m_h$ peak coincides 
with the large $|A_0|$ bulge of Fig. \ref{fig:m0mhf}{\it d}) which leads
to large stop mixing. As is well known, large stop mixing leads to large 
radiative corrections to $m_h$\cite{mhiggs,mhiggs2,h125} and lifts $m_h$ up into
the $\sim 125$ GeV range. 
This is shown in a scatter plot of events with appropriate EWSB and $m_Z^{PU}<4m_Z^{OU}$ 
in the $m_h$ vs. $A_0$ plane in Fig.~\ref{fig:mhvsA0}. There we see explicitly that
large $-A_0$ events correlate with large $m_h\simeq 125$ GeV while lower values of
$-A_0$ correspond to too low a value of $m_h$. 
This is a testable prediction of the 
string landscape picture: a value of $m_h\sim 125$ GeV is reflective of 
large stop mixing which can be untangled for instance at an $e^+e^-$ collider
operating with $\sqrt{s}> 2m_{\tst_1}$\cite{Moortgat-Picka:2015yla}. 
Other stringy scenarios-- such as G2MSSM\cite{Kane:2011kj} 
or mini-split SUSY\cite{Arvanitaki:2012ps,ArkaniHamed:2012gw}--
obtain $m_h\sim 125$ GeV via very heavy (unnatural) top squarks but 
with rather small stop mixing. The low $m_h\sim 120$ GeV bump comes from
small stop mixing, with $-A_0\alt 1$ TeV.

\begin{figure}[htb]
\begin{center}
\includegraphics[height=0.4\textheight]{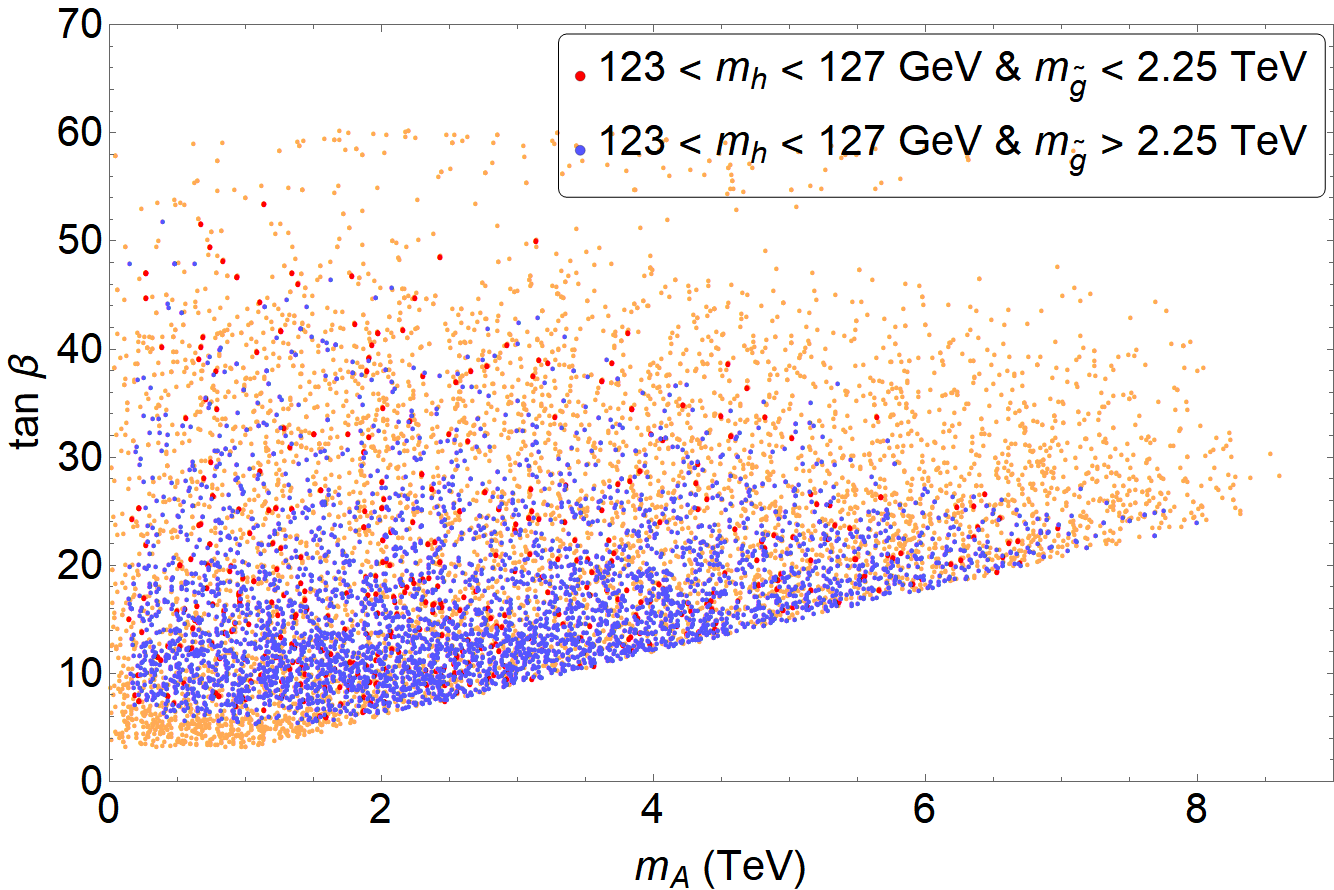}
\caption{Scatter plot of models with appropriate EWSB and $m_Z^{PU}<4m_Z^{OU}$
in the $\tan\beta$ vs. $m_A$ plane with $\mu =150$ GeV. 
The orange dots have $m_h<123$ GeV.
\label{fig:tanbvmA}}
\end{center}
\end{figure}

In Fig. \ref{fig:higgs}{\it b}), the distribution in $dP/dm_A$ peaks around 
$m_A\sim 1.5-2$ TeV with a long tail extending as high as 8 TeV. The upper
limit on $m_A$ arises from the $m_{H_d}^2/\tan^2\beta$ contribution
to the weak scale in Eq. \ref{eq:mzs}. For $\tan\beta\sim 10$, then 
$m_A\alt 3$ TeV but for larger values of  $\tan\beta$, 
then $m_A\sim m_{H_d}$ can become much bigger: see the $\tan\beta$ vs. $m_A$
scatter plot of solutions in Fig. \ref{fig:tanbvmA}.
Such large values of $m_A\gg m_h$ 
(and consequently $m_H$ and $m_{H^\pm}$) predict that the Higgs sector looks
decoupled, with $h$ behaving largely as a SM-like Higgs boson. 
Thus, the landscape SUSY prediction is that precision Higgs measurements at 
HL-LHC or an $e^+e^-$ Higgs factory will see at best only small deviations 
from SM Higgs properties.

In frame {\it c}), we see the distribution in $\tan\beta$, 
which is scanned uniformly since it is not a soft term. 
The expected value is for $\tan\beta\sim 10-20$ with a falling tail 
towards larger values. Large $\tan\beta$ yields large bottom-squark
$\Sigma_u^u(\tb_{1,2})$ and tau-lepton contributions  to the weak scale 
and so is disfavored. 

\begin{figure}[H]
\begin{center}
\includegraphics[height=0.39\textwidth]{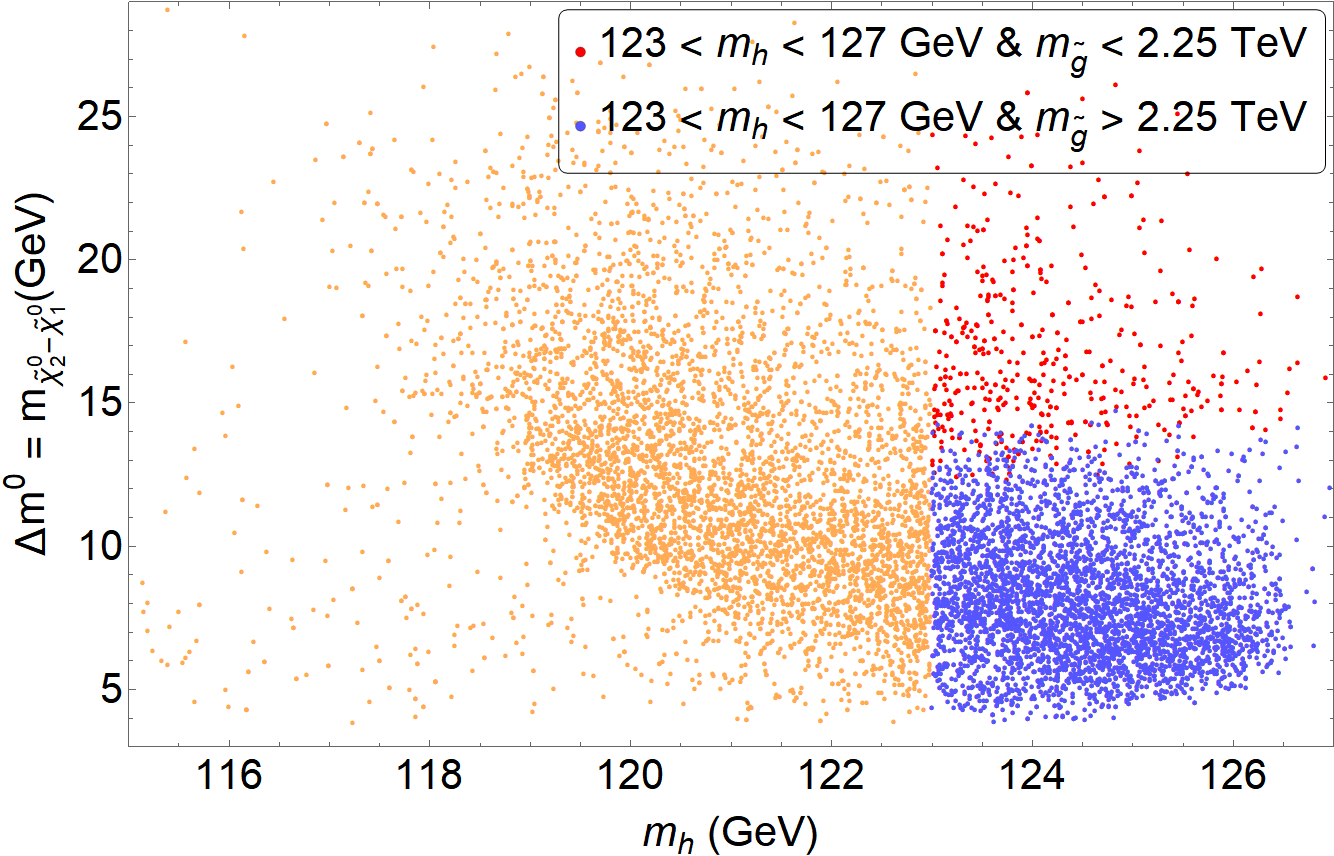}
\caption{Scatter plot of models with appropriate EWSB and $m_Z^{PU}<4m_Z^{OU}$
in the $\Delta m^0$ vs. $m_h$ plane with $\mu =150$ GeV. 
The orange dots have $m_h<123$ GeV.
\label{fig:mhvsDm0}}
\end{center}
\end{figure}

In frame {\it d}), we show the light Higgsino mass
gap $\Delta m^0\equiv m_{\tz_2}-m_{\tz_1}$. 
This quantity is relevant to light higgsino
pair production at LHC via the soft-opposite-sign-dilepton plus jet plus MET
channel\cite{Baer:2011ec,Han:2014kaa,Baer:2014kya,Han:2015lma,Baer:2016usl,Baer:2020sgm,CMS:2016zvj,Aad:2019qnd}. 
The prediction here is for a mass gap $\Delta m^0\sim 8-12$ GeV but with a tail
extending to $15-25$ GeV. 
However, the large $\Delta m^0>14$ GeV tail corresponds to models with $m_h<123$ GeV 
and $m_{\tg}\alt 2.25$ TeV
as can be seen from the scatter plot of $\Delta m^0$ vs. $m_h$ in Fig. \ref{fig:mhvsDm0}.
In comparison, the $n=0$ draw predicts much larger
mass gaps while $n=1$ or 2 predicts gaps typically below 10 GeV.
Such mass gaps can be easily measured to high precision at an
$e^+e^-$ collider operating with $\sqrt{s}>2 m(higgsino)$\cite{ilc,Baer:2019gvu}.

\begin{figure}[H]
\begin{center}
\includegraphics[height=0.22\textheight]{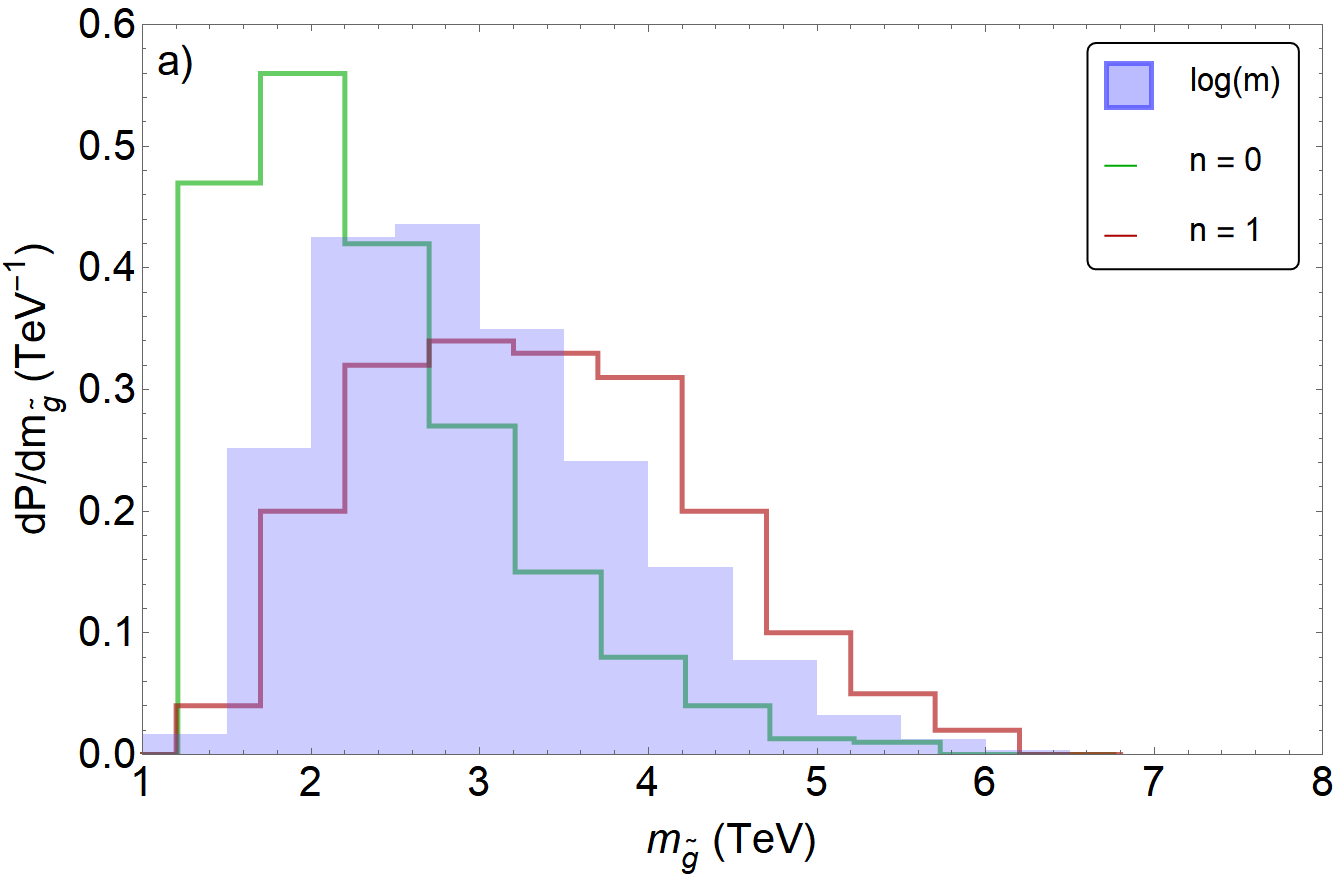}
\includegraphics[height=0.22\textheight]{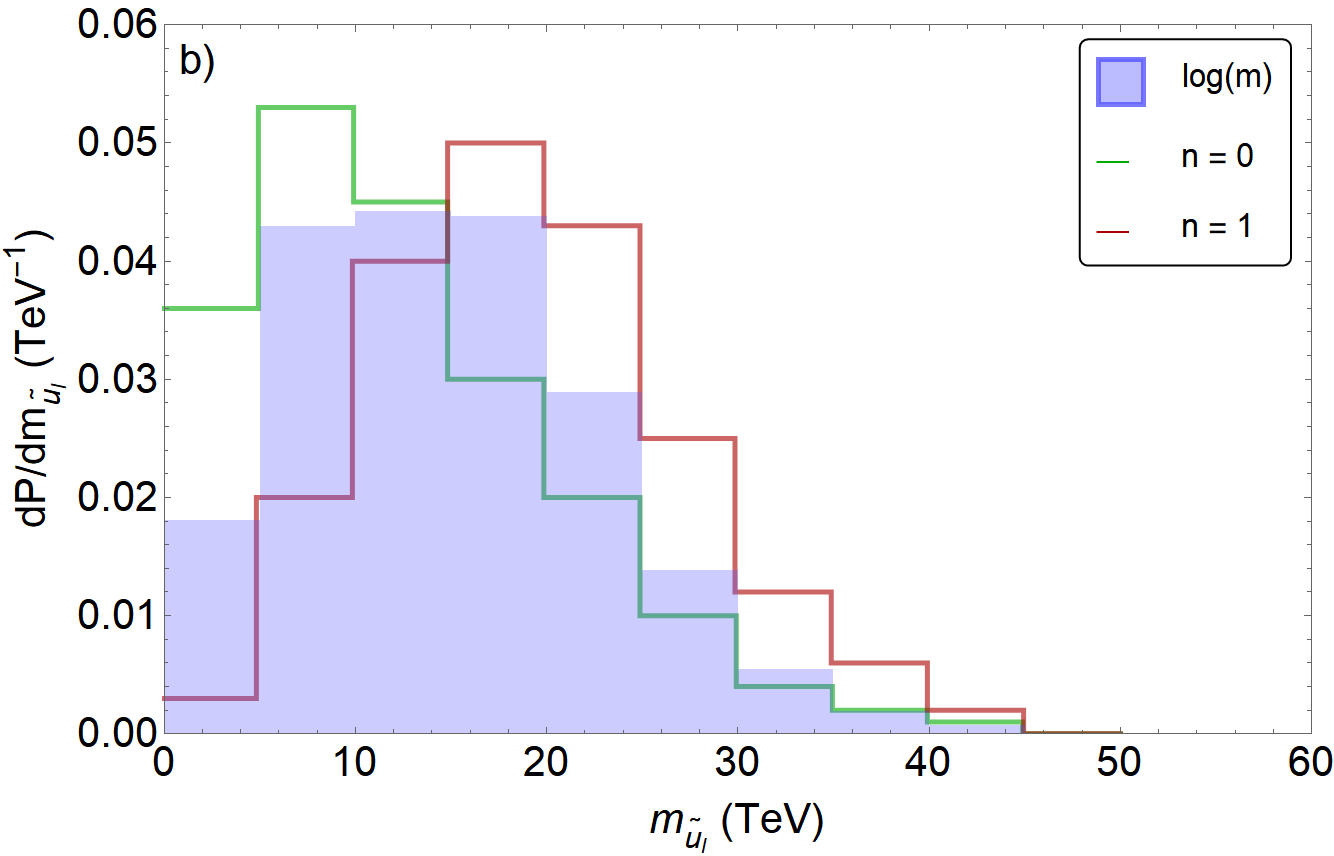}\\
\includegraphics[height=0.22\textheight]{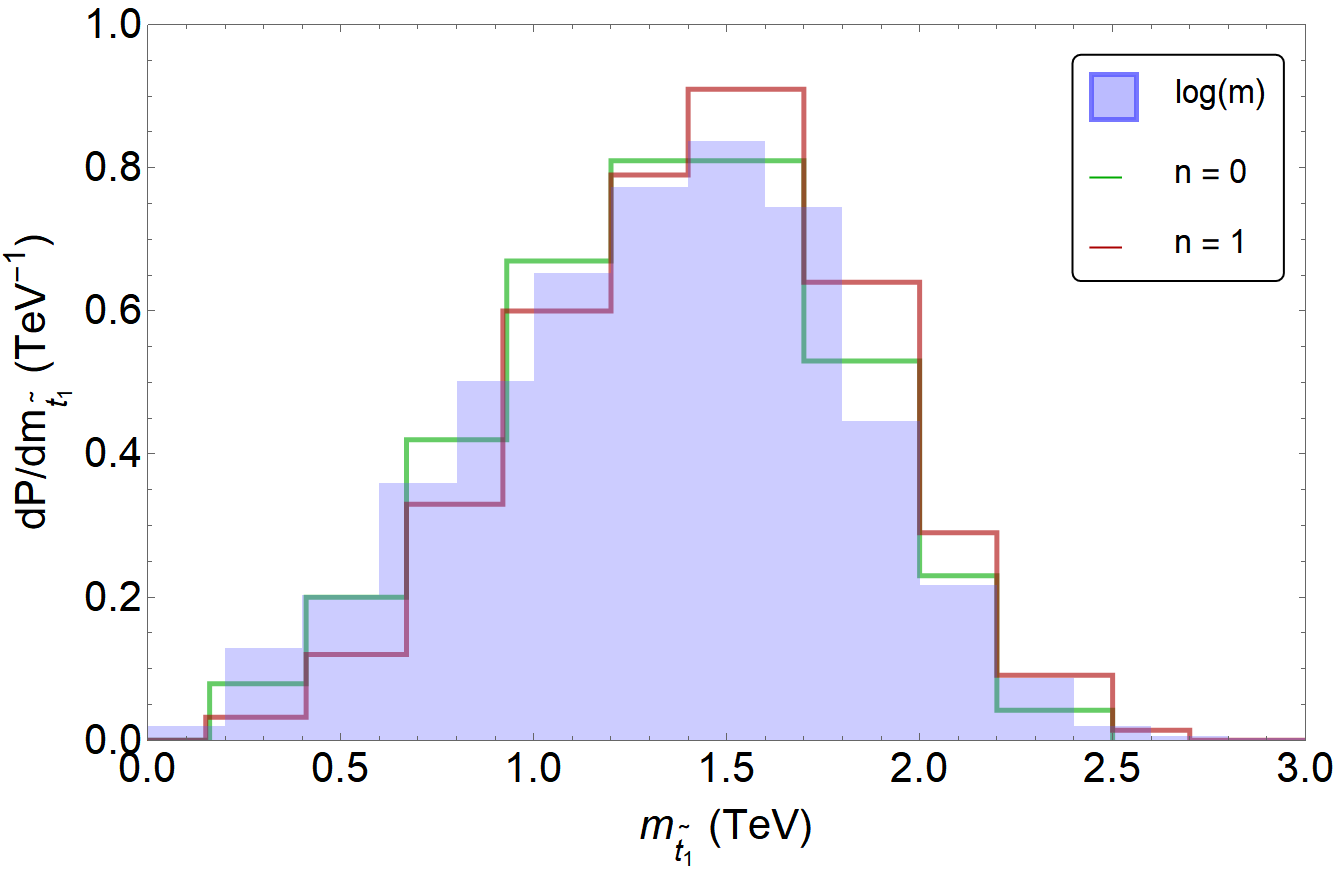}
\includegraphics[height=0.22\textheight]{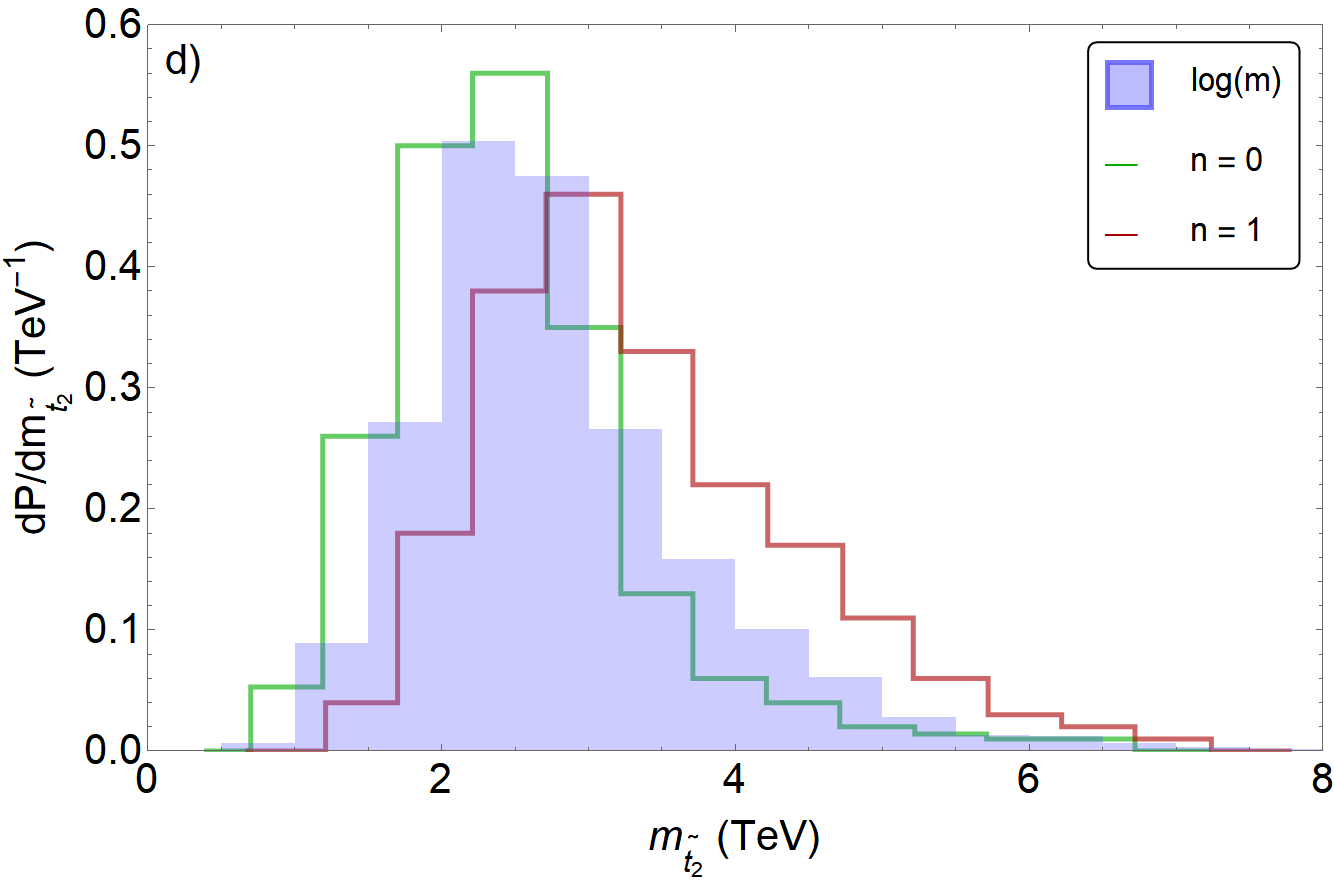}
\caption{Probability distributions for NUHM3 masses and 
parameters {\it a}) $m_{\tg}$, {\it b}) $m_{\tu_L}$, {\it c}) $m_{\tst_1}$ and
{\it d}) $m_{\tst_2}$ from a log distribution of soft terms
in the string landscape with $\mu =150$ GeV.
For comparison, we also show probability distributions for $m_{soft}^n$ with $n=0$ and 1.
\label{fig:glsq}}
\end{center}
\end{figure}

In Fig. \ref{fig:glsq}, we show the predicted probability distributions for
various strongly interacting sparticles which are relevant for LHC
SUSY searches. In frame {\it a}), we show the distribution 
$dP/dm_{\tg}$. The log distribution has a peak at $m_{\tg}\sim 2.5-3$ TeV, 
somewhat beyond present LHC limits which require $m_{\tg}\agt 2.2$ TeV.
Only a small portion of log parameter space is excluded by the present LHC 
gluino mass limit, so from the landscape point of view, it is not surprising
that LHC has not discovered SUSY. The distributions tails off to values
of $m_{\tg}\sim 5-6$ TeV-- such high values of $m_{\tg}$ would require
at least an energy doubling upgrade of LHC for detection\cite{Baer:2017pba,Baer:2018hpb}. 
In frame {\it b}), we show an example of first/second generation squark masses: 
in this case $m_{\tu_L}$. Here we see a peak value of $m_{\tu_L}\sim 10-20$ TeV
with a tail extending to $\sim 40$ TeV. Such high first/second generation 
scalar masses are pulled upward until their two-loop RGE contributions to
the top-squark sector cause those soft terms to run tachyonic resulting 
in CCB vacua.

In frame {\it c}), we show $dP/dm_{\tst_1}$. 
For the log distribution, the peak is around $m_{\tst_1}\sim 1.5$ TeV. 
This compares to current LHC limits which require $m_{\tst_1}\agt 1.1$ TeV. 
Thus, again we see that it is 
not surprising that top-squarks have not been detected at LHC from 
the string landscape point of view. In frame {\it d}), we show the distribution
$dP/dm_{\tst_2}$. This distribution peaks at $m_{\tst_2}\sim 3$ TeV and would 
likely require at least an energy doubling of LHC in order to gain 
detection\cite{Baer:2017pba,Baer:2018hpb}. The log distribution is again
intermediate beween the $n=0$ and $n=1$ scan results.

\section{Conclusions}
\label{sec:conclude}

In this paper, we have been motivated by two major success stories for
the string theory landscape: 1. its success in explaining the tiny value of
the cosmological constant and 2. when applied to the statistics of SUSY
breaking with a power-law draw to large soft terms and an anthropic bound on the
pocket universe value of the weak scale $m_{weak}^{PU}$, it predicts a
value $m_h\simeq 125$ GeV with sparticles (other than hard-to-detect higgsinos) 
lifted beyond present LHC search limits. In this paper, we were motivated
by an earlier suggestion by Dine based on DSB that soft terms would instead
have a logarithmic statistical distribution. This case has also been recently
advanced by Broeckel {\it et al.}\cite{Broeckel:2020fdz} where they consider the effects of 
K\"ahler moduli stabilization to derive a log distribution for soft terms
from models of LVS stabilization. Alternatively, they find a power-law
soft term statistical distribution for the KKLT and PS models.

The three scenarios are listed in Table \ref{tab:models} along with
the expected form of soft terms: either from gravity mediation or mirage
mediation. Higgs and sparticle mass distributions have previously been
presented in Ref. \cite{Baer:2017uvn} for power-law draw and gravity mediation
as expected in PS. In Ref. \cite{Baer:2019tee}, the corresponding
distributions have been presented for power-law draw but with mirage mediation.
In this work, we complete the Table by presenting expected Higgs and sparticle
statistical mass distributions for a log draw with gravity mediation, 
as may be expected from the LVS stabilization within the string landscape.
\begin{table}[H]
\renewcommand{\arraystretch}{1.0}
\begin{center}
\begin{tabular}{|c|ccc|}
\hline
model & KKLT\cite{Kachru:2003aw} & LVS\cite{Balasubramanian:2005zx} & PS\cite{Berg:2005yu} \\
\hline
soft terms & mirage & grav & grav \\
\hline
soft dist. & $m_{soft}^n$ & $\log (m_{soft})$ & $m_{soft}^n$\\
\hline
mass dist'ns & \cite{Baer:2019tee} & this paper & \cite{Baer:2017uvn} \\
\hline
\end{tabular}
\caption{Three models of moduli stabilization along with expected form
of soft terms, expected soft term distribution in string IIB landscape
and reference for associated statistical distributions of 
Higgs and sparticle masses.
}
\label{tab:models}
\end{center}
\end{table} 

Our results from Fig. \ref{fig:higgs} show that the expectation for
$m_h\simeq 125$ GeV still holds for a log draw, although the Higgs peak at 
125 GeV is not as sharp as that found from $n=1$ or 2 power-law draw.
Also, sparticles (other than higgsinos) are still expected to lie beyond
LHC search limits, but again not as sharply as in the power-law case.
Of course, if experiment confirms SUSY particles within the ranges shown here, 
one will be able to distinguish mirage from gravity-mediation, for instance, 
by extracting running gaugino masses at an $e^+e^-$ collider and checking 
at which scale they might unify\cite{ilc2}. But for the case of gravity mediation, 
based on just a single data point of SUSY spectra, there will be no way to 
distinguish between the different stabilization mechanisms LVS or PS.

{\it Acknowledgements:} 
We thank Kuver Sinha for discussions.
We thank Dr. Hasan Serce for help on increasing statistics in the updated version.
This material is based upon work supported by the U.S. Department of Energy, 
Office of Science, Office of High Energy Physics under Award Number DE-SC-0009956. 
The computing for this project was performed at the OU Supercomputing Center 
for Education \& Research (OSCER) at the University of Oklahoma (OU).


\bibliography{logland2}
\bibliographystyle{elsarticle-num}

\end{document}